\documentclass{optica-article}

\usepackage{silence}
\journal{opticajournal} % for journals or Optica Open

\articletype{Research Article}

\usepackage{lineno}
\usepackage{bm}
\begin{document}

\title{Scalable All-Optical Fibre-Mode Data Transmission with Profiles-Preserved Decoding}

\author{Rundong Fan,\authormark{1,2,3,4}, Yamin Zheng,\authormark{4}, Pei Li,\authormark{1,2,3}, Xixiao Cao,\authormark{1,3,6}, Haoyang Wu,\authormark{1,2,3}, Zheng Cai,\authormark{4,5} and Lei Huang,\authormark{1,2,3,4,5*}}

\address{\authormark{1}Key Laboratory of Photonic Control Technology (Tsinghua University), Ministry of Education, Beijing 100084, China\\
\authormark{2}State Key Laboratory of Precision Space-time Information Sensing Technology, Beijing 100084, China\\
\authormark{3}Department of Precision Instrument, Tsinghua University, Beijing 100084, China\\
\authormark{4}Department of Astronomy, Tsinghua University, Beijing 100084, China\\
\authormark{5}Deep Space Technology Center, Tsinghua University, Beijing 100084, China\\
\authormark{6}Weiyang College, Tsinghua University, Beijing 100084, China}

\email{\authormark{*}hl@tsinghua.edu.cn} %% email address is required; see note below about the corresponding author designation

% use {asbstract*} to suppress the copyright line. Copyright information will be added in production

\begin{abstract*} 
Optical fibres are the primary medium for optical signal transmission, and their guided modes provide a high-dimensional basis for modal-domain information encoding. However, conventional demultiplexing approaches typically convert fibre modes into fundamental Gaussian modes and require repeated mode conversions, while existing profiles-preserved methods are generally restricted to fewer than three modes. High-quality fibre-mode data transmission therefore requires a scalable all-optical decoder capable of separating strongly overlapping modal channels while preserving their intrinsic spatial profiles. Here, we establish a scalable profiles-preserved all-optical decoding method for high-dimensional fibre-mode data transmission. By introducing a microlens-array-assisted decoding architecture with channel-dependent spherical phase compensation, the proposed method accommodates mode-dependent effective focal-length variations, enabling scalable modal-channel separation while preserving high-quality modal profiles at the output plane. Experimentally, the optical decoder resolved fields containing eight fibre modes, achieving a mode fidelity exceeding 0.72, a worst-channel crosstalk of $-5.57~\mathrm{dB}$ and a mean non-target crosstalk of $-21.34~\mathrm{dB}$, while reconstructing the relative modal weights with an error below 0.1. Semantic transmission experiments using digits and Chinese characters further demonstrated effective recovery of the encoded information from the decoded modal signals. We expect this work to provide a scalable route towards high-dimensional all-optical fibre-mode data transmission.
\end{abstract*}

%%%%%%%%%%%%%%%%%%%%%%%%%%  body  %%%%%%%%%%%%%%%%%%%%%%%%%%
\section{Introduction}
Optical signal transmission is widely used in modern communication systems owing to its high speed and low loss. As the fundamental medium for optical signal delivery, optical fibres have undergone rapid development driven by the growing demand for data capacity. To address this demand, multiplexing techniques that exploit different degrees of freedom of light, such as wavelength, phase, amplitude and spatial mode, have been widely developed to enable parallel transmission. Among them, wavelength-division multiplexing (WDM) \cite{liu2024parallel, luo2013time, shoeib2024fiber, hao2023non, erkilincc2017bidirectional} and space-division multiplexing (SDM) \cite{van2014ultra, zou2022high, puttnam2021space, pauwels2019space, murshid2008spatial} represent two major routes for capacity scaling. In SDM, the guided modes supported by few-mode and multimode fibres provide a high-dimensional basis for modal-domain information encoding and parallel transmission. Increasing the core diameter of an optical fibre allows it to support a larger number of guided modes, thereby enhancing its potential information-carrying capacity. However, when a fibre supports a large number of guided modes, often on the order of hundreds, strong intermodal coupling can occur during propagation, leading to modal crosstalk and increased transmission loss. These effects complicate signal recovery and limit practical transmission distances \cite{richardson2013space, ploschner2015seeing}.

Mode-division multiplexing (MDM), as a major branch of SDM, exploits orthogonal fibre modes as parallel data channels to increase information capacity. In such systems, the optical field received after propagation is generally a superposition of multiple modal components, and the transmitted data can only be recovered by identifying the constituent modes and extracting their corresponding modal weights. Therefore, the mode demultiplexer functions as an optical decoder at the receiver, directly determining the scalability, crosstalk, and reconstruction fidelity of fibre-mode data transmission. Conventional optical decoders include Mach–Zehnder interferometers (MZIs) \cite{ohta2018si, igarashi2015selective}, photonic lanterns (PLs) \cite{velazquez2018scaling, dana2024free} and silicon photonic waveguides (SPWs) \cite{luo2014wdm, stern2015chip}, which typically involve trade-offs among the number of accessible modes, device footprint, and intermodal crosstalk. Recent micro-optical and nanophotonic devices have further expanded modal decoding capabilities. For example, multi-plane light conversion has been used to map arrays of fundamental fibre modes onto higher-order Laguerre–Gaussian modes, enabling the conversion of hundreds of modes \cite{fontaine2019laguerre}. The reverse process has also been demonstrated using multi-plane microstructures, where a set of fibre modes (such as $\mathrm{LP}_{01}$, $\mathrm{LP}_{11a}$ and $\mathrm{LP}_{11b}$) is mapped into arrays of fundamental modes, enabling ultra-compact and low-crosstalk mode demultiplexing \cite{oh2022adjoint}. Overall, most existing modal decoders rely on conversion-based strategies that map fibre modes into spatially separated fundamental-mode Gaussian spots.

Although conversion-based modal decoding is effective for channel separation, it does not preserve the intrinsic spatial profiles of individual fibre modes. In long-haul optical systems employing mode multiplexing, subsequent processes such as signal amplification\cite{marhic2015fiber}, modulation\cite{lu2021spatial} and re-coupling may require mode-selective operations\cite{gong2016all, zhao2024design}, for which preserving the original modal profiles is highly desirable. This has motivated increasing interest in profiles-preserved mode demultiplexing, which enables spatial separation while retaining the intrinsic field distributions of individual modes. For example, a profiles-preserved mode-demultiplexing metalens has been proposed that exploits differences in the spatial field distributions of fibre modes \cite{xu2025metasurface}. By superimposing spherical phase profiles with different focal lengths at distinct spatial locations, the metalens functions analogously to a microlens array and can spatially demultiplex the $\mathrm{LP}_{01}$, $\mathrm{LP}_{11a}$ and $\mathrm{LP}_{11b}$ modes while preserving their intrinsic spatial profiles, thereby eliminating additional mode-conversion processes during transmission. However, existing profiles-preserved approaches are generally limited to only a few low-order modes. As the mode order and modal dimensionality increase, different fibre modes exhibit progressively stronger spatial overlap, making region-modulation-based demultiplexing schemes difficult to extend to high-order and many-mode scenarios. Therefore, scalable profiles-preserved optical decoding of strongly overlapping fibre modes is essential for advancing high-dimensional fibre-mode data transmission.

To overcome this scalability limitation, the decoder must implement a trainable and globally coupled spatial-field transformation rather than relying solely on local region modulation. The profiles-preserved separation of strongly overlapping fibre modes can therefore be viewed as a modal feature extraction and remapping process, in which the mixed input field is optically transformed into spatially separated output channels while retaining the intrinsic modal profiles. Optical neural networks (ONNs) provide a natural platform for such all-optical transformations, as computation is performed through wave propagation, diffraction and coherent interference across multiple trainable phase layers. Owing to their intrinsic parallelism, high speed and low power consumption, ONNs have recently been demonstrated in applications including pulse shaping \cite{veli2021terahertz, zhou2024spatiotemporal, zhang2024space}, information encryption \cite{park2026recomposable, guo2025polarization, lei2025diffractive} and orbital angular momentum (OAM) beam demultiplexing \cite{feng2025high, bozinovic2013terabit, yan2014high}. These properties, combined with prior physical knowledge of fibre modes, make ONNs highly suitable for scalable profiles-preserved all-optical decoding of high-dimensional fibre-mode data.

In this study, we establish a scalable profiles-preserved all-optical decoding method for high-dimensional fibre-mode data transmission. The method is implemented by an optical modal decoder that performs the key demultiplexing and profile-preserving transformation in the optical domain. As illustrated in Fig. 1(a), the decoder receives an overlapping fibre modal field and maps its constituent modal components to distinct output positions while preserving their intrinsic spatial profiles. The decoding function is achieved through a sequence of trainable phase modulation layers. In particular, microlens-array-assisted spherical phase compensation is incorporated into the first and final layers to mitigate the influence of mode-dependent variations in effective focal length. This design enables profiles-preserved decoding beyond the few-mode regime and provides a route towards scalable modal-channel separation. We experimentally validate the proposed method at both the modal-decoding and data-transmission levels. First, an eight-mode optical decoder was demonstrated, which can separate and identify individual modal channels from overlapping incident fields. The worst-channel crosstalk and the mean non-target crosstalk were as low as $-5.57~\mathrm{dB}$ and $-21.34~\mathrm{dB}$, respectively, while the profiles-preserved modal fidelity exceeded 0.72. The relative modal weights are further reconstructed from the separated output patterns, confirming the capability of modal-content recovery. The proposed framework can, in principle, be extended to a larger number of modes or other modal bases. To verify its use in fibre-mode data transmission, we further encode digits and Chinese characters into modal-domain binary signals using an eight-channel mode-multiplexing scheme. At the receiver, the encoded information is recovered through overlap-integral-based coefficient extraction and threshold decoding, and the reconstructed semantic information is fully consistent with the original input. Collectively, these results demonstrate the potential of scalable profiles-preserved all-optical decoding for high-dimensional fibre-mode data transmission.

\section{Result}
\subsection{Principle of scalable profiles-preserved all-optical decoding}
To implement profiles-preserved all-optical decoding, the proposed modal decoder uses multilayer diffractive modulation to perform trainable modal filtering and channel remapping in the optical domain. As shown in Fig. 1(b), the operating principle of the DNN is based on the Huygens–Fresnel principle \cite{depasse1995huygens}. As the optical field propagates through a diffractive layer, each pixel acts as a secondary wave source, emitting spherical secondary waves that coherently interact with all pixels in the subsequent layer. The amplitude and phase of these secondary waves are jointly determined by the incident optical field and the complex-valued transmission (or reflection) coefficient imposed at each pixel. Through the cumulative wave interference across multiple diffractive layers, the DNN realizes a globally coupled wave-based transformation of the input optical field. Detailed theoretical derivations and implementation details are provided in Note S1 (Supporting Information).

During fibre-mode decoding, the primary function of the optical decoder is to selectively extract the modal component associated with a target channel from an overlapping incident field, map it onto a predefined output position and suppress contributions from non-target modes. The field emerging from the fibre end face is a divergent beam with spherical phase components. From the perspective of Fourier optics \cite{goodman1969introduction}, different fibre modes can be interpreted as having distinct spatial-frequency characteristics, so the mode-screening process can be regarded as spatial-frequency filtering of modal components followed by spatial remapping to the corresponding output channel. As shown in Fig.~1(c), a transmission-type architecture was adopted in the DNN decoder design. For a single fibre mode propagating with a spherical wavefront, a compensating spherical phase with focal length $f$ is introduced at the first diffractive layer, and the propagation distance from the fibre end face to this layer is also set to $f$. This operation converts the divergent modal field into a spatial-domain distribution suitable for subsequent optical filtering. After modulation and propagation through the following diffractive layers, an additional spherical phase is applied at the final diffractive layer to re-map the filtered field to the output plane. In the decoder design, the parameter $f$ is treated as a trainable variable, enabling adaptive matching to the equivalent focal lengths of different fibre modes. The theoretical background of fibre modes and detailed results for single-mode decoding using the proposed optical decoder are provided in Notes~S2 and S3 of the Supplementary Information.

It has been reported that different fibre modes exhibit spherical wavefronts with distinct equivalent focal lengths \cite{xu2025metasurface}. Therefore, designing a multimode optical decoder based on a single spherical-wave compensation may lead to blurred output profiles, analogous to defocus in conventional optical systems. To address this issue, we introduce microlens-array phase modulation into the diffractive decoding architecture, as shown in Fig.~1(c). Specifically, microlens phase arrays with adjustable focal lengths are incorporated into the first and final diffractive layers. As the coupled optical field propagates in free space, it gradually expands to cover the entire microlens-array region. Since each spatial region of the coupled fibre field contains information from the supported modes, the first microlens array partitions the optical field into different modal-decoding channels and applies divergent spherical-wave phase compensation with distinct focal lengths. This enables channel-specific spatial-frequency mapping for different fibre modes. Subsequently, based on the spatial-frequency filtering process described above, the target modal component is extracted within each channel. Finally, a convergent spherical-wave phase is introduced in the last diffractive layer to transform the filtered distribution back into the spatial-domain modal profile. In this way, the proposed decoder enables scalable modal-channel separation while preserving the intrinsic spatial profiles of the fibre modes.

The focal lengths of the microlenses vary across different lateral positions. Furthermore, the centre positions and number of microlenses are mapped one-to-one to the spatial positions and number of modes, respectively. The DNN is optimized iteratively using adaptive moment estimation (Adam) \cite{amari1993backpropagation, tian2023recent}, with forward propagation modeled by the Rayleigh–Sommerfeld method \cite{gao2019rayleigh}. For the theoretical details of the Adam optimisation method, please refer to Note S1.3 (Supporting Information). The output field is segmented according to predefined pattern-space positions ($x_i$, $y_i$), and the overlap integral $C_i$ with the interpolated and scaled ideal LP mode profiles is computed \cite{ploschner2015seeing}, as shown in Fig. 1(c). The loss function is then defined as:
\begin{equation}
\mathrm{loss}(\bm{f}, \bm{L}, \bm{\varphi})
=
\sqrt{
\frac{
\sum_{i=1}^{M}
\left(
C_i - \hat{C}_i / a
\right)^2
}{M}
}
\end{equation}
\begin{equation}
C_i
=
\iint_{\Omega}
\sqrt{I_i(x,y)}
\cdot
\left| P_i^{*}(x,y) \right|
\, dx \, dy ,
\end{equation}
where, $\hat{C}_i$ denotes the LP mode coefficients of the training dataset, which serve as supervisory labels. The $a$ is a parameter greater than 1 that controls the energy characteristics during DNN design. Because the network’s function is to extract valid patterns while discarding invalid ones, the transmission matrix of the ONN is non-unitary \cite{macho2021optical, reck1994experimental}, resulting in inherent energy loss. During training, adjusting this parameter allows the network to better manage energy transmission, ensuring that all neurons can fully contribute to the pattern extraction function. where, $I_i$ is the intensity distribution of the $i^{\mathrm{th}}$ output channel, $P_i$ is the corresponding field distribution of the $i^{\mathrm{th}}$ LP mode, and $\Omega$ denotes the integration region of the segmented image. The ''*'' indicates the complex conjugate operation on a matrix. During ONN training, the trainable parameters are $(\bm{f}, \bm{L}, \bm{\varphi})$, where $\bm{f}$ denotes the spherical focal lengths of each microlens in the first and last diffraction layers, with $\bm{f}=(f_1^1,f_1^2,\ldots,f_1^M,f_2^1,f_2^2,\ldots,f_2^M)$, and $\bm{\varphi}=(\bm{\varphi}_1,\bm{\varphi}_2,\ldots,\bm{\varphi}_N)$ represents the phase profile of each diffraction layer. Here, $i=1,2,\ldots,M$, where $M$ is the number of analyzed patterns and $N$ is the total number of layers in the ONN. $\bm{L}=(L_1,L_2)$ denotes the propagation distances from the input optical field to the ONN and from the ONN to the output plane, respectively.

\subsection{Eight-channel profiles-preserved modal decoding}
Figure~2(a) illustrates the experimental setup used to validate the eight-channel profiles-preserved optical modal decoder. 
\begin{figure}[htbp]
\centering\includegraphics[width=12.5cm]{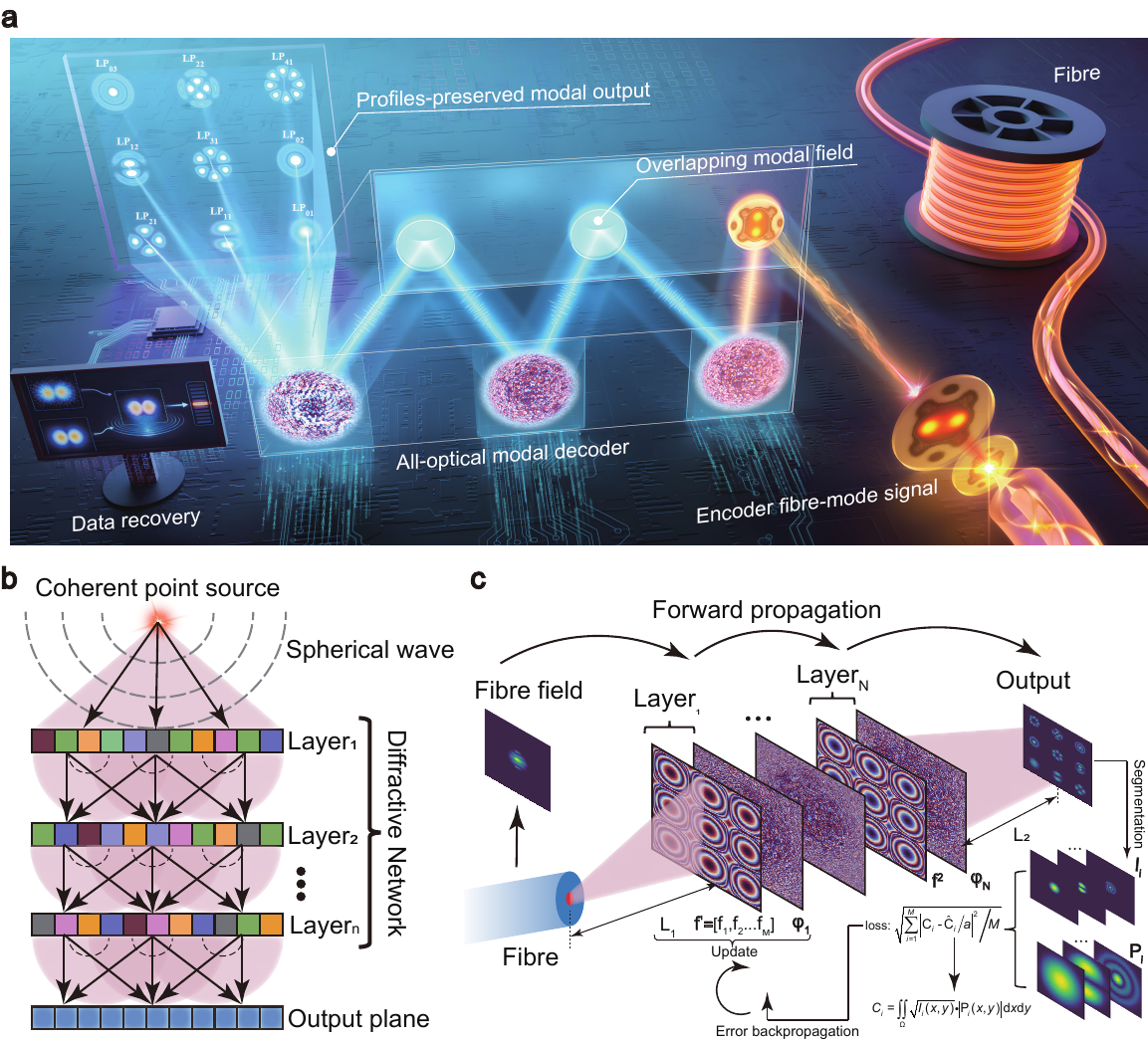}
\caption{Principle of scalable profiles-preserved all-optical decoding for fibre-mode data transmission. (a) Schematic illustration of the proposed all-optical modal decoding process. An overlapping fibre modal field carrying multiple modal channels is spatially separated by the optical decoder into distinct output positions while preserving the intrinsic spatial profiles of the corresponding modes, such as $\mathrm{LP}_{01}$, $\mathrm{LP}_{11}$ and $\mathrm{LP}_{22}$. (b) Operating principle of the diffractive optical decoder. Secondary waves generated by the diffractive units in the preceding layer coherently interact with subsequent diffractive layers, resulting in a globally coupled wave-propagation process for trainable modal filtering and channel remapping. (c) Design principle of the profiles-preserved optical modal decoder. Microlens-array phase compensation is introduced to account for the mode-dependent effective focal lengths of different modal channels, while overlap integration is employed for modal-coefficient extraction and loss-function construction during iterative optimisation.}
\end{figure}
In the experiment, a laser source with a wavelength of $780~\mathrm{nm}$ was employed. The system consists of two modules: modal-field encoding and all-optical modal decoding. In the encoding module, the first spatial light modulator, $\mathrm{SLM}_1$, was used to generate optical field distributions corresponding to high-order fibre modes. This strategy avoids the limitation of conventional mode converters, such as PLs and MZIs, which are generally restricted to the generation of low-order fibre modes. The generated optical field was filtered through a $4f$ system and imaged onto $\mathrm{CCD}_1$ for observation of the modal intensity distributions. Detailed methods for generating optical fields using $\mathrm{SLMs}$ \cite{maurer2011spatial, lazarev2012lcos} are provided in Note~S4 of the Supporting Information. The conjugated optical field from the $4f$ system was guided into the modal-decoding path through a non-polarising beam splitter (NPBS) and a planar mirror. The all-optical decoding function was implemented through alternating reflections between the planar reflector and $\mathrm{SLM}_2$, and the resulting output field was imaged onto $\mathrm{CCD}_2$ through another $4f$ system. Detailed information on the experimental setup is provided in the Methods section. The optimised phase distributions of the eight-channel optical decoder are shown in Fig.~2(b). The decoder consists of three diffractive layers, each with a resolution of $900 \times 900$ pixels. In addition, a blazed grating phase was incorporated into the background of the decoded hologram to suppress stray light caused by misalignment of the diffractive layers.
\begin{figure}[htbp]
\centering\includegraphics[width=12cm]{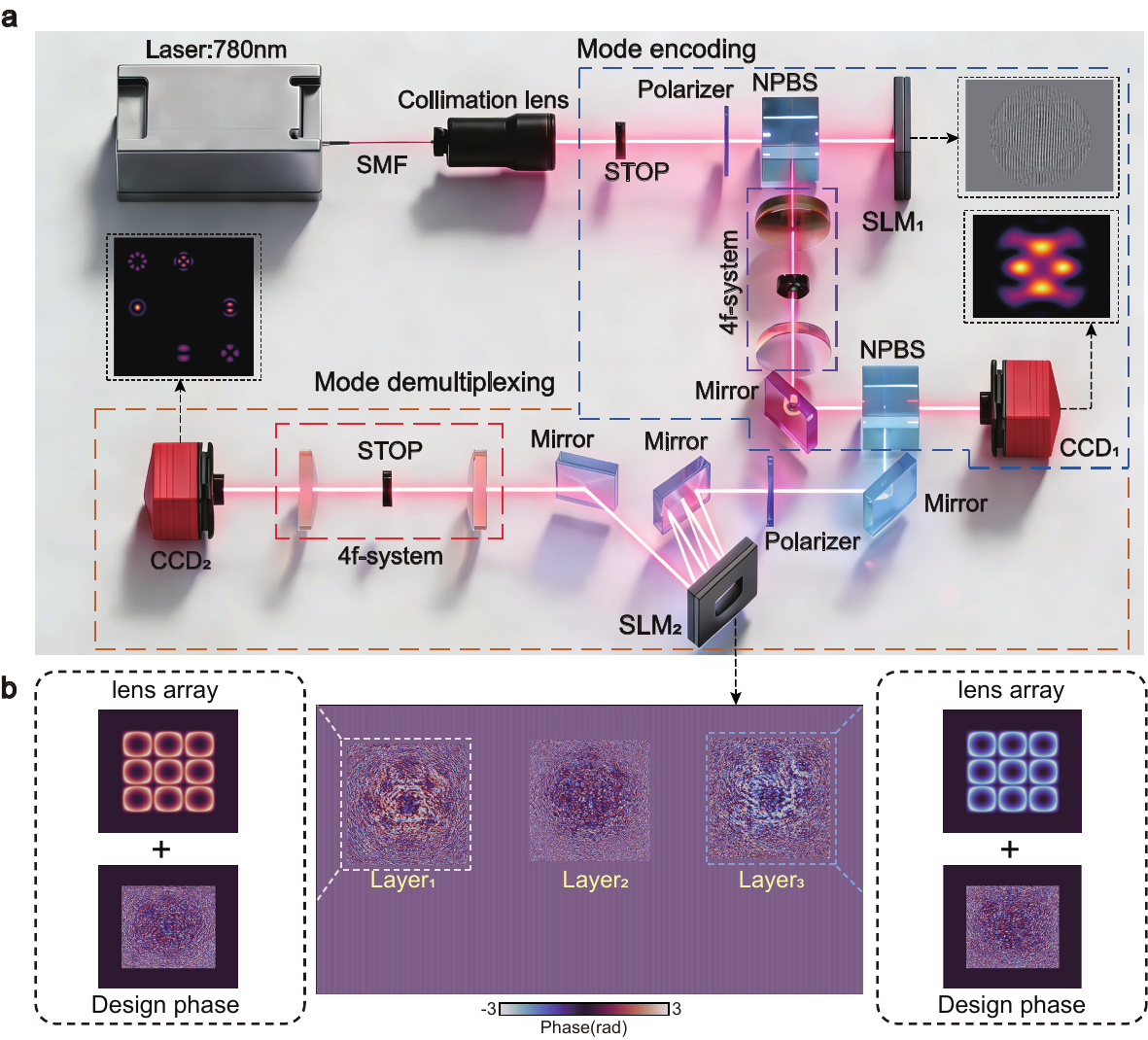}
\caption{Experimental setup for eight-channel profiles-preserved all-optical modal decoding. (a) Experimental optical layout of the multi-plane reflective modal-decoding system. The setup consists of two modules: a modal-field encoding path, which encodes a fundamental-mode Gaussian beam into an optical field containing high-order or mixed fibre modes, and an all-optical modal-decoding path, which spatially separates the overlapping modal components while preserving their profiles. Both encoding and decoding are implemented using spatial light modulators ($\mathrm{SLMs}$). PC: personal computer. (b) Full holographic phase maps of the eight-channel optical modal decoder. The decoder consists of three holographic phase maps, each incorporating pixel-to-pixel alignment compensation.}
\end{figure} 

In the experiment, the optical field was incident on $\mathrm{SLM}_2$ at an oblique angle. Previous studies \cite{zhang2023multi} have shown that when the incident angle is below $5^\circ$, the propagation of the incident field can still be accurately approximated using the normal-incidence transmission model. Accordingly, the incident angle in our optical system was set to $4.8^\circ$, representing a compromise between optical alignment constraints and experimental performance. Specifically, because the lateral positions of the diffractive layers on the $\mathrm{SLM}$ are fixed, a larger incident angle reduces the available spacing between $\mathrm{SLM}_2$ and the reflector, whereas a smaller incident angle increases the likelihood of beam blockage by optical components. Here, the distance between $\mathrm{SLM}_2$ and the mirror was $30~\mathrm{mm}$, and the mirror had a lateral size of $12.5~\mathrm{mm} \times 12.5~\mathrm{mm}$. In addition, oblique incidence introduces projection distortion, leading to an effective shift in the pixel distribution on the $\mathrm{SLM}$. To mitigate this effect, a superpixel strategy was adopted in the decoder design, where $3 \times 3$ pixels were combined into a single superpixel. The optical decoder was therefore initially designed at a resolution of $300 \times 300$ and subsequently expanded to $900 \times 900$. Detailed discussions are provided in Note~S4.2 of the Supporting Information.

First, $\mathrm{SLM}_1$ was used to generate all target single-mode patterns, including $\mathrm{LP}_{01}$, $\mathrm{LP}_{11}$ and higher-order $\mathrm{LP}_{22}$ modes. Each mode was incident onto the optical decoder, and the corresponding output distribution was recorded. In the experiment, the input beam diameter on the decoder was $2~\mathrm{mm}$. Figure~3(a) shows the intensity distributions of the decoded modes. The first column displays the simulated results under single-mode incidence, while the second and third columns show the experimental results and the corresponding magnified views, respectively. The observed output positions indicate that the designed diffractive layers can accurately map different modes to their predefined spatial positions. The crosstalk between modal channels directly determines whether each mode can serve as an independent data-carrying channel, and therefore represents a key factor limiting the capacity of the mode-multiplexing system. Figure~3(b) shows the inter-channel crosstalk of the proposed optical decoder. Each channel maintains a low crosstalk level, with the maximum crosstalk among the eight channels being $-5.57~\mathrm{dB}$. For the non-target modal channels, the maximum crosstalk is $-6.06~\mathrm{dB}$ and the mean value is $-21.34~\mathrm{dB}$. These results indicate that the decoded modal channels can be used as independent channels for data transmission. To further quantitatively evaluate the profiles-preserved decoding fidelity, the structural similarity index (SSIM) between each reconstructed modal distribution and the corresponding ideal modal intensity profile was calculated, as shown in Fig.~3(c). The results show that the proposed optical decoder can spatially separate different modal channels while retaining their intrinsic spatial profiles, thereby maintaining high fidelity within the designed operating range. The average SSIM exceeds 0.72, with a maximum value of 0.86 and a minimum value of 0.595, corresponding to the $\mathrm{LP}_{22}$ mode. To the best of our knowledge, the proposed approach represents one of the most scalable reported solutions for profiles-preserved optical mode multiplexing, supporting a larger number of modes while maintaining a low crosstalk level\cite{xu2025metasurface, zhao2025neuro}.

Furthermore, the optical decoder was experimentally evaluated for modal-weight reconstruction using two-mode mixed fields with different relative weights. The intensity distributions at the corresponding output positions were analysed to determine whether they accurately reflected the modal content of the incident field. In this experiment, mixed optical fields consisting of $0.5\times\mathrm{LP}_{01}+1\times\mathrm{LP}_{11}$, $1\times\mathrm{LP}_{01}+0.5\times\mathrm{LP}_{11}$ and $1\times\mathrm{LP}_{01}+1\times\mathrm{LP}_{11}$ were incident onto the decoder. The corresponding input field distributions are provided in Fig.~S9 of the Supporting Information. Subsequently, overlap integration was used to extract the relative modal weights of $\mathrm{LP}_{01}$ and $\mathrm{LP}_{11}$ from the output results, as shown in Fig.~3(d). It should be noted that the calculated values represent relative modal weights rather than absolute modal powers, since the absolute modal content is influenced by the incident laser power. The experimental results demonstrate that the modal content of the incident optical field can be effectively reconstructed by calculating the overlap integral between each output pattern and the corresponding ideal modal profile. Among the three groups of measurements, the maximum root-mean-square (RMS) reconstruction error was 0.083, while the minimum error was 0.006. As shown in Figs.~3(e)--3(g), modes with lower relative weights correspond to weaker output intensity distributions. These results confirm that the proposed optical decoder can simultaneously manipulate multiple fibre modes, including high-order modal channels. It recovers the relative modal-channel information while preserving the intrinsic spatial profile of each mode. Meanwhile, the inter-channel crosstalk between different modal channels is suppressed to a relatively low level, enabling the decoded patterns to function as independent parallel channels. This capability provides a promising approach for realizing large-capacity fibre-mode data transmission systems.

\subsection{Fibre-mode data transmission of semantic information}
In the previous section, we verified that the proposed optical decoder can perform profiles-preserved modal-channel separation and relative modal-weight reconstruction from overlapping incident fields. These two capabilities provide the basis for fibre-mode data transmission, in which information is encoded into modal-domain states and recovered by extracting the corresponding modal coefficients at the receiver. In principle, the proposed decoding framework can be extended to a larger number of fibre modes. In practice, however, its performance is constrained by the alignment accuracy and modulation capability of the experimental system. As the number of supported modes increases and the modal distributions become more complex, higher alignment precision is required. Nevertheless, compared with conventional approaches, the proposed optical decoder enables the manipulation and separation of a larger number of high-order fibre modes, thereby increasing the available modal channels for information transmission. Based on this modal-domain transmission scheme, we conducted semantic information transmission experiments as a representative fibre-mode data-transmission task. As illustrated in Fig.~4(a), the original semantic signal was first converted into a binary sequence, and each bit was assigned to one fibre-mode channel, where the presence and absence of a given mode represented the binary states ''1'' and ''0'', respectively. The resulting multi-mode superposition formed the transmitted optical field. At the receiver, the proposed all-optical decoder separated the modal channels, after which overlap-integral-based coefficient extraction and threshold decision were used to recover the binary sequence and identify the transmitted semantic information.

\begin{figure}[htbp]
\centering\includegraphics[width=12.0cm]{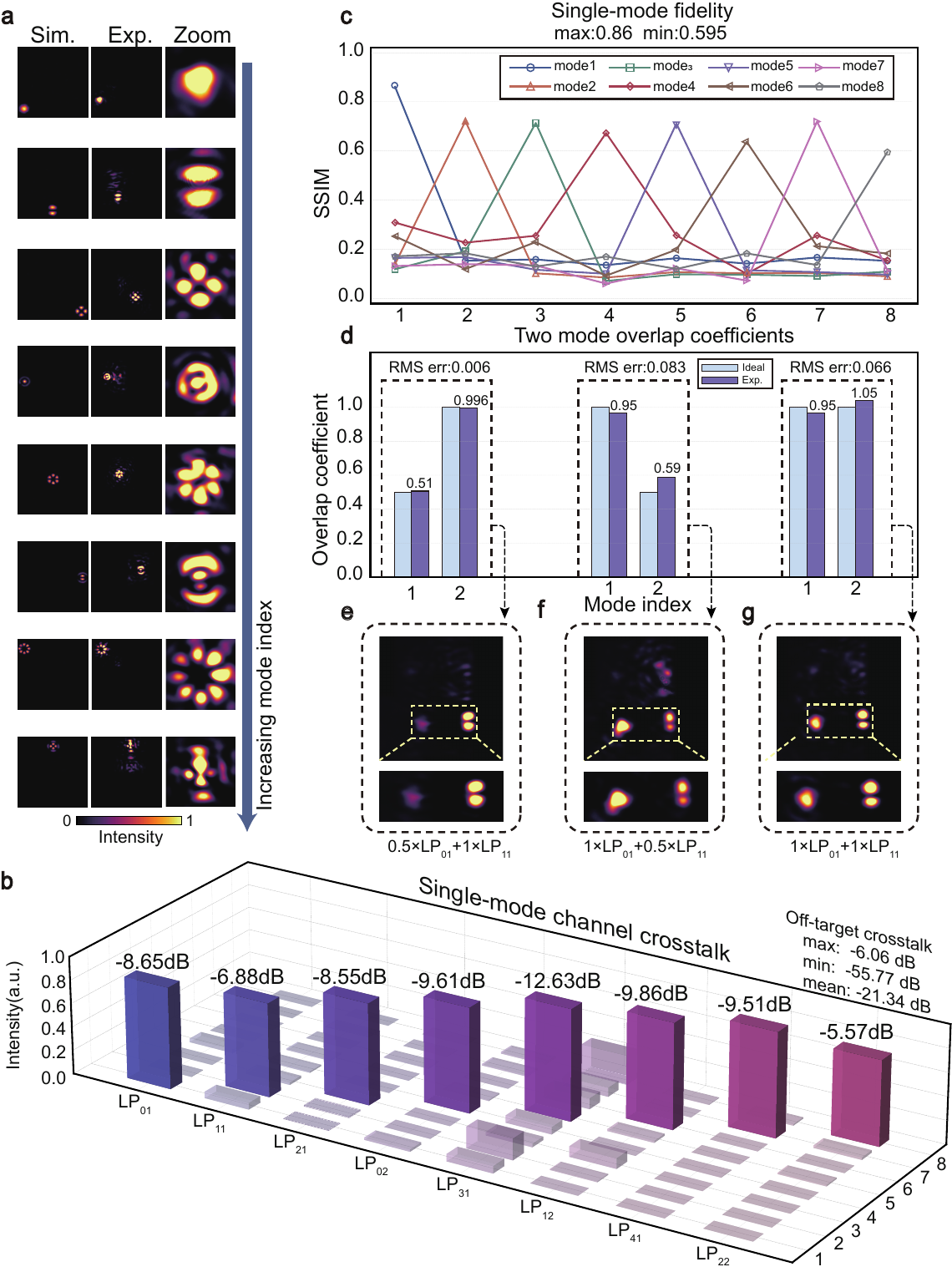}
\caption{Experimental results of the eight-channel profiles-preserved optical modal decoder. 
(a) Simulated results, experimental results and magnified experimental images under single-mode incidence, shown from left to right. 
(d) Crosstalk analysis of the optical decoder. The crosstalk values of individual modal channels are labelled above the corresponding output channels.
(c) Single-mode decoding fidelity of the optical decoder, quantified by the similarity between the output intensity distribution at the corresponding spatial position and the ideal modal distribution. 
(d) Reconstructed modal-weight histograms obtained from the optical decoder under incident fields with different relative modal weights. The corresponding values represent the relative weights of the incident modes. 
(e)--(g) Output intensity distributions corresponding to three incident fields with different relative modal weights. These results correspond to the modal-weight reconstructions shown in (d).}
\end{figure}

In the experiments, the transmitted semantic information consisted of ''1911'' and the Chinese characters ''Tsing'' and ''Hua''. Since both the numerical and Chinese-character information exceeded the encoding capacity of 8 bits, each signal was encoded into a 16-bit binary sequence and divided into two groups of 8-bit modal distributions. In this manner, each semantic signal was represented by two modal distribution maps, enabling a more comprehensive evaluation of the modal-decoding and information-recovery capability of the proposed system. The encoder used in the experiment consisted of a personal computer (PC) and an SLM. The encoding strategy for the semantic signals and the corresponding incident modal distributions are shown in Fig.~S8 of the Supporting Information. Based on the recovered binary sequences, the original semantic information was successfully restored.

The simulated and experimental intensity distributions of the six groups of mixed-mode fields corresponding to the three semantic signals are shown in Figs.~4(b) and 4(c), respectively. The experimental results demonstrate that the proposed all-optical decoder can separate the encoded modal channels, map them to the designated spatial positions and preserve their spatial profiles. From the perspective of fibre-mode data transmission, this process converts a mixed modal field carrying multiple binary states into spatially resolved modal outputs, allowing the transmitted signal to be recovered through direct coefficient extraction rather than additional mode-conversion stages. This confirms that modal-domain signals can be optically decoded from mixed-mode input fields. Stray light can also be observed in the experimental output distributions. This effect may originate from two main factors. First, as the number and order of encoded modes increase, the system becomes more sensitive to lateral misalignment between diffractive layers, resulting in stray light caused by assembly errors. Second, the limited effective resolution of the $\mathrm{SLM}$ used for mode encoding introduces distortions in the generated incident modal distributions, thereby reducing the modal-recognition accuracy of the decoder. The alignment sensitivity of the system could potentially be alleviated by integrating multilayer microstructures onto the fibre endface through lithographic fabrication processes~\cite{yu2025all,principe2017optical,schmidt2020tailored}, which have been extensively investigated in recent years. In this work, a pixel-to-pixel alignment strategy was adopted to compensate for lateral shifts between diffractive layers in the holographic implementation. Detailed alignment procedures and the holograms used for calibration are provided in Note~S5 of the Supporting Information. Subsequently, the modal weights in the output images were calculated using the overlap-integration method, as shown in Fig.~4(d). \begin{figure}[htbp]
\centering\includegraphics[width=13.0cm]{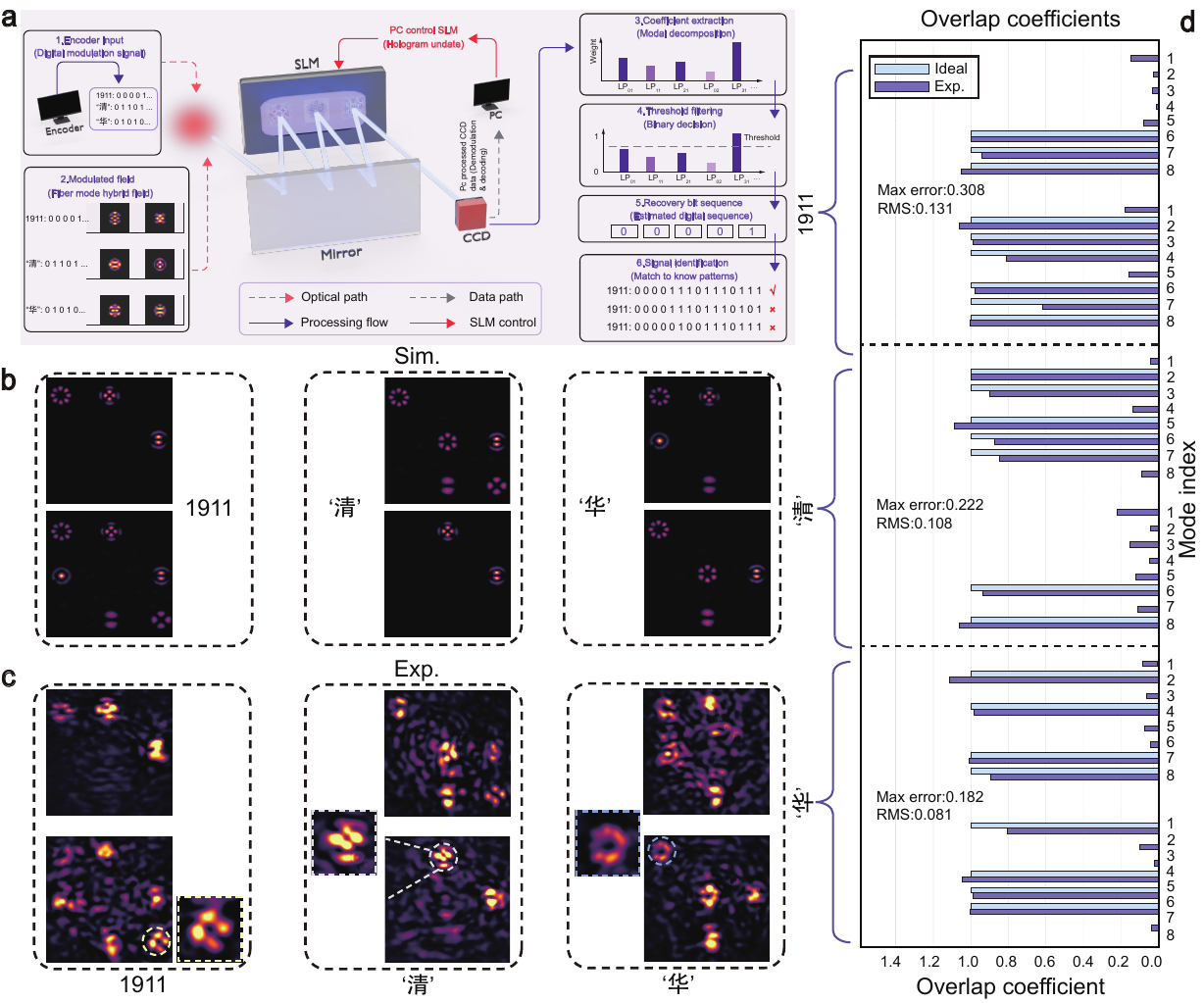}
\caption{Experimental results of semantic information transmission using fibre-mode encoding. 
(a) Conceptual illustration of modal-domain signal transmission and recovery based on the proposed all-optical decoder. 
(b) Simulated intensity distributions of the decoded semantic signals, corresponding from left to right to ''1911'', ''Tsing'' and ''Hua''. 
(c) Experimental intensity distributions of the decoded semantic signals. The experimental results exhibit a one-to-one correspondence with the simulated results shown in (b). 
(d) Overlap-integration results calculated from the experimental intensity distributions in (c), corresponding from top to bottom to ''1911'', ''Tsing'' and ''Hua''.}
\end{figure}
The results demonstrate that the dominant modal components of the incident optical fields were successfully reconstructed, enabling recovery of the transmitted binary signals. The maximum relative weight errors for the three semantic signals were 0.308, 0.222 and 0.182, respectively, while the corresponding RMS errors were 0.131, 0.108 and 0.081. The largest reconstruction error occurred for the ''1911'' signal, owing to its larger number of encoded modal patterns and the inclusion of multiple high-order modes. As shown in Fig.~4(d), although reconstruction errors caused by stray light are present, the recovered relative modal weights can still be reliably converted into binary data through thresholding. In this experiment, a threshold value of 0.4 enabled accurate reconstruction of the transmitted information.

Furthermore, to further evaluate the proposed scheme under a more realistic fibre-mode transmission scenario, qualitative experiments were conducted using real few-mode optical fibres. In these experiments, we investigated whether the proposed optical decoder could separate fibre modes with known modal compositions from experimentally generated fibre outputs, and whether significant modal crosstalk or degradation effects appeared after propagation and decoding. The corresponding experimental results are provided in Note~S6 of the Supporting Information. These results further confirm the feasibility of using the proposed profiles-preserved all-optical decoder for modal-channel recovery in practical fibre systems. By converting overlapping fibre-mode fields into spatially resolved and profile-preserved modal outputs, the proposed approach enables direct modal-coefficient extraction for information recovery, avoiding repeated conversions between fundamental Gaussian modes and high-order fibre modes. This capability is particularly important for fibre-mode data transmission, as it allows the modal signals carried by individual high-order modes to be preserved after demultiplexing, thereby facilitating subsequent mode-selective processing and re-coupling.

%文章书签2026.5.25
\section{Discussion}
Overall, we have established a scalable profiles-preserved all-optical decoding method for high-dimensional fibre-mode data transmission. The proposed optical decoder spatially separates overlapping fibre-mode components while preserving their intrinsic modal profiles, enabling direct modal-coefficient extraction for information recovery. By introducing a microlens-array-assisted phase architecture to compensate for the mode-dependent defocus among different fibre modes, the system achieves accurate and scalable modal-channel separation. The reconstructed modal profiles achieved a fidelity exceeding 0.72. In addition, the proposed decoder responds to different relative modal weights in the incident optical field, enabling effective recovery of modal coefficients through overlap integration. We further demonstrated its use in semantic information transmission. By encoding the transmitted information into 16-bit binary data and mapping it into the modal space, followed by all-optical decoding and threshold decision, the transmitted information was accurately reconstructed, with a maximum relative-weight error of 0.308. Importantly, the proposed decoding architecture can, in principle, be extended to larger numbers of modes or other modal bases, thereby addressing the scalability limitations of existing profiles-preserved approaches. By preserving the modal signals carried by individual high-order modes after demultiplexing, this method may facilitate subsequent mode-selective processing and re-coupling, providing a potential route towards large-capacity fibre-mode data transmission with reduced mode-conversion complexity. We expect this work to contribute to the development of next-generation multimodal photonic information processing and optical communication systems.

In fibre-based communication systems, system integration is of critical importance. In this work, we further validated the feasibility of integrating scalable profiles-preserved all-optical decoding with fibre-mode data transmission. However, the current implementation remains constrained by alignment accuracy and device miniaturisation. Future work will focus on introducing cascaded metastructures into the proposed optical decoding framework, with the aim of realizing compact and integrated high-order fibre-mode decoding for large-capacity modal-domain transmission \cite{georgi2021optical, mei2023cascaded}.  In addition, the proposed approach may provide a route towards overcoming the dimensional limitations of conventional fibre-mode transmission systems. By incorporating additional optical degrees of freedom, such as wavelength and polarisation, into the modal-domain encoding and decoding framework, the proposed architecture could potentially support more versatile and higher-dimensional optical communication schemes, thereby expanding the practical application range of fibre-mode data transmission systems.
\begin{backmatter}
\bmsection{Methods}
\noindent\textbf{Laboratory procedure.} The experimental setup is shown in Fig.~2(a). The overall optical system consists of two modules: optical field encoding and mode demultiplexing. These two functions are implemented using separate spatial light modulators (SLMs). In the optical field encoding path, a checkerboard phase modulation method was employed to achieve complex-amplitude modulation using a single SLM. This approach enables the transformation of a Gaussian beam into arbitrary higher-order or mode-mixed optical fields. The generated intensity distributions of individual modes and mixed modes are presented in Figs.~S9 and S10 of the Supporting Information. The SLM used for mode encoding was model UPOLabs HDSLM80R, with a resolution of $1920 \times 1200$ and a pixel pitch of $8~\mu\mathrm{m}$. Complex-amplitude modulation of coherent optical fields using the checkerboard phase modulation technique inevitably introduces a significant amount of stray light. To suppress this effect, a 4f system was employed to filter the effective optical field. In the experiment, the \(+1\) diffraction order was extracted by superimposing a blazed grating phase onto the hologram, with a grating period of 10 pixels.
The $4f$ system consisted of two lenses with focal lengths of \(f = 300~\mathrm{mm}\) and \(f = 225~\mathrm{mm}\), respectively. The difference in focal lengths enabled scaling of the optical field, allowing a larger beam size to illuminate the SLM and thereby improving the effective modulation resolution. In the experiment, the diameter of the collimated beam incident on the SLM was 3~mm. Before entering the mode demultiplexing path, a polariser was used to eliminate possible changes in the optical field introduced during the front-end modulation process.

In the mode demultiplexing path, multiple holographic phase distributions needed to be displayed simultaneously on a single SLM. Therefore, a large-area 4K-resolution SLM was employed. The SLM used in the experiment was the UPOLabs HDSLM38R, featuring a resolution of \(4096 \times 2160\) pixels and a pixel pitch of \(3.8~\mu\mathrm{m}\). After modulation by the SLM, another $4f$ optical system was employed to scale the optical field for observation. In this optical path, the two lenses used had focal lengths of \(f = 200~\mathrm{mm}\) and \(f = 225~\mathrm{mm}\), respectively.

\noindent\textbf{Multi-plane Reflective Optical Path Alignment.} The primary challenge in the multi-plane reflective optical system lies in the alignment between diffractive layers. In this work, the alignment procedure was divided into two stages. The first stage involved alignment of the SLM and the planar reflector to ensure approximate parallelism. During this process, a pinhole-based alignment method was employed for positioning and calibration. A five-axis adjustment stage was mounted on the SLM, and the planar reflector was used as the alignment reference. The SLM orientation was then adjusted such that the reflected beam propagated through the same pinhole position.

 The second stage involved alignment of the diffractive holograms. In this process, a pixel-compensation strategy was adopted by laterally shifting the holograms displayed on the SLM to compensate for offsets introduced by residual angular misalignment of the optical system. The hologram alignment procedure consisted of both coarse and fine adjustment steps. These processes were implemented using two sets of dedicated holographic patterns, as detailed in Note~S5 of the Supporting Information.

\bmsection{Funding}
National Key Research and Development Program of China, 2023YFA1605600; Tsinghua University Initiative Scientific Research Program; Tsinghua University Education Foundation; Tsinghua-Jiangyin Innovation Special Fund (TJISF).

% \bmsection{Acknowledgments}
% The section title should not follow the numbering scheme of the body of the paper. Additional information crediting individuals who contributed to the work being reported, clarifying who received funding from a particular source, or other information that does not fit the criteria for the funding block may also be included; for example, ``K. Flockhart thanks the National Science Foundation for help identifying collaborators for this work.'' 

\bmsection{Disclosures}
The authors declare no conflicts of interest.

\bmsection{Data Availability Statement}
All phase patterns used for optical fibre mode generation and demultiplexing have been made publicly available in image format. These phase patterns are compatible with mainstream \(\mathrm{SLM}\) platforms, provided that the resolution and pixel pitch specifications described in the main text are satisfied. All ONN models were trained using the PyTorch framework. The pretrained models for single-mode extraction, multi-mode extraction, and ONN-based semantic transmission are provided in \texttt{.pth} format. All pretrained models supporting the findings of this study are available at \url{https://github.com/THUAO-Lab/DNN-PPD}.

\bmsection{Code availability}
All codes used for the design and simulation in this work have been made openly available. The open-source framework was developed for the design of diffractive neural networks for conformal fibre mode decomposition. The Python-based code supports flexible adjustment of the number of diffractive layers and superpixel dimensions, and has been accelerated using CUDA computing. In addition, the modelling and simulation tools for fibre modes have also been publicly released. The open-source repository can be accessed at \url{https://github.com/THUAO-Lab/DNN-PPD}. Notably, the proposed framework is not limited to fibre modes and can be readily extended to other modal bases, such as Hermite–Gaussian modes.
\end{backmatter}

%%%%%%%%%%%%%%%%%%%%%%% References %%%%%%%%%%%%%%%%%%%%%%%%%

%%%%%%%%%% If using BibTeX:
\bibliography{sample}

%%%%%%%%%% If preparing manually:
% \begin{thebibliography}{1}
% \newcommand{\enquote}[1]{``#1''}

% \bibitem{Zhang:14}
% Y.~Zhang, S.~Qiao, L.~Sun, Q.~W. Shi, W.~Huang, L.~Li, and Z.~Yang,
%   \enquote{Photoinduced active terahertz metamaterials with nanostructured
%   vanadium dioxide film deposited by sol-gel method,}
%   {\protect\JournalTitle{Optics Express}} \textbf{22}, 11070--11078 (2014).

% \bibitem{Optica}
% {Optica}, \enquote{{Optica Publishing Group},}
%   \url{http://www.opg.optica.org}.

% \bibitem{FORSTER2007}
% P.~Forster, V.~Ramaswamy, P.~Artaxo, T.~Bernsten, R.~Betts, D.~Fahey,
%   J.~Haywood, J.~Lean, D.~Lowe, G.~Myhre, J.~Nganga, R.~Prinn, G.~Raga,
%   M.~Schulz, and R.~V. Dorland, \enquote{Changes in atmospheric consituents and
%   in radiative forcing,} in \enquote{Climate Change 2007: The Physical Science
%   Basis. Contribution of Working Group 1 to the Fourth Assesment Report of
%   Intergovernmental Panel on Climate Change,}  S.~Solomon, D.~Qin, M.~Manning,
%   Z.~Chen, M.~Marquis, K.~B. Averyt, M.~Tignor, and H.~L. Miler, eds.
%   (Cambridge University Press, 2007).

% \end{thebibliography}

\end{document}